# Magnetic Correlations In Deuteronium Jarosite, A Model S = 5/2 Kagomé Antiferromagnet

A. S. Wills and A. Harrison*,

*Department of Chemistry, University of Edinburgh, The King's Buildings, West Mains Rd., Edinburgh, EH9 3JJ, UK.*

S. A. M. Mentink and T. E. Mason

*Department of Physics, University of Toronto, 60 St. George Street, Toronto, Ontario, M5S 1A7, Canada.*

Z. Tun

*Atomic Energy of Canada Limited, Chalk River, Ontario, Canada, K0J 1JO*

* Author to whom correspondence should be addressed

Deuteronium jarosite, $(D_3O)Fe_3(SO_4)_2(OD)_6$, contains a kagomé lattice of Heisenberg spins S=5/2 with a coverage of 97±1%. DC and AC susceptibility measurements show strong in-plane antiferromagnetic exchange ($\theta_{CW}$ = –1500±300K) and a spin-glass transition at $T_f$ = 13.8K, while the magnetic contribution to the specific heat below $T_f$ rises with T as $T^2$, characteristic of two-dimensional propagating modes. Powder neutron diffraction reveals short-range magnetic correlations ($\xi \approx$ 19±2 Å) with a wavevector corresponding to the $\sqrt{3} \times \sqrt{3}$ spin structure at 1.9K.

The Heisenberg kagomé antiferromagnet presents a fascinating challenge for theorists[1-3] and experimentalists[4-8] alike. The lattice is built from vertex-sharing triangles in which the frustration of nearest-neighbour antiferromagnetic exchange produces a 120° structure within each triangle (Figs 1(a) and 1(b)). For coplanar spin configurations a chirality (handedness) $n$ may be associated with each triangular plaquette, where $n = (2/3\sqrt{3})\sum_{i,j}(S_i \times S_j)$, and $S_i$, $S_j$ are the spins taken in a clockwise sense around the plaquette. Neither the 120° spin structure nor the chirality is conveyed uniquely between triangles through a shared vertex[9], resulting in a highly degenerate ground state. Thermal and quantum fluctuations lift this degeneracy, selecting spin configurations with softer, more highly entropic excitations. This "order-by-disorder" process selects coplanar configurations in the limit $T \to 0$, two of which are shown in Figs 1(a) and (b), and referred to as the q = 0 and the √3 x √3 structure respectively. Both these spin structures can support cooperative fluctuations of blocks of spins and the creation of extended spin defects bounded by spins of one sublattice only [2]. Moments within the blocks may be rotated by an arbitrary angle $\phi$ about the direction of the peripheral spins at no energy cost, creating open and closed spin folds for the q = 0 and √3 x √3 structures respectively. Although it takes no energy to create a single open or closed $\phi = \pi$ fold in a coplanar spin state, interactions between such defects give rise to glassy magnetic behaviour. It is still not clear what is the most appropriate description of the spin correlations in the medium that supports these defects. The √3 x √3 structure is weakly favoured over the q = 0 structure, but Monte Carlo simulations indicate that the appropriate order parameter does not saturate at very low temperatures, and no order is observed in the staggered chirality that would be expected to accompany it, leading to doubt as to whether conventional long-range order can occur as $T \to 0$.

Theoretical work has been complemented, and in many cases stimulated, by experimental work on the layered garnet $SrGa_{12-x}Cr_xO_{19}$ (SCGO(x)) in which Heisenberg spins S=3/2 sit at the vertices of a kagomé lattice[4] coupled by strong antiferromagnetic exchange (the Curie-Weiss constant $\theta_{CW} = -500K$). DC magnetic susceptibility ($\chi_{dc}$) measurements reveal a spin-glass transition at $T_f = 3-7K$, depending on the composition $x$; this is accompanied by a progressive slowing-down of the fluctuations as SCGO(x) is cooled from well above $T_f$ [5] However, this type of spin-glass is most peculiar. Elastic and inelastic neutron scattering data show short-range correlations corresponding to the $\sqrt{3} \times \sqrt{3}$ structure with a correlation length of $\approx 7\text{Å}$ (approximately twice the Cr-Cr separation)[6], though *ca.* 2/3 of the scattering at $T = 0.5T_f$ is from excitations whose energy is greater than 50GHz, and no static component of the moment could be observed down to 0.01K by muon spin relaxation measurements for which the time-scale of the probe is 170MHz[5]. Specific heat measurements reveal an unusual $T^2$ dependence below $T_f$[7] normally associated with long-range two-dimensional magnetic order.

Unfortunately, SCGO(x) has shortcomings as a model kagomé antiferromagnet. The coverage of the kagomé lattice is far from complete, being typically 89 - 92%, though dilution studies indicate that the glassy behaviour is characteristic of the perfectly covered lattice. Further, a significant proportion of the $Cr^{3+}$ ions reside in triangular lattices between the kagomé layers, complicating analysis of the magnetic properties. Recent measurements of the spin excitations in SCGO(x) have shown that this material is best described as a layer of corner sharing tetrahedra (made up of 2 kagomé layers and an intervening triangular layer) separated by triangular layers of spins bound in singlets[8]. Thus, while it is a highly frustrated system, SCGO(x) is far from a faithful representation of the simple kagomé lattice which has been the subject of most theoretical work. In this letter we discuss such a system in the form of a deuterated analogue of the mineral hydronium iron jarosite, $(H_3O)Fe_3(SO_4)_2(OH)_6$, in which the coverage of the kagomé lattice is 97±1%. Measurements of $\chi_{dc}$ and the AC susceptibility ($\chi_{ac}$) reveal strong antiferromagnetic exchange ($\theta_{CW} = -1500\pm300$K) and a spin-glass transition at $T_f = 13.8$K (from $\chi_{dc}$), while neutron scattering measurements demonstrate that the magnetic correlations remain short-range down to 1.9K, with a wavevector corresponding to the $\sqrt{3} \times \sqrt{3}$ structure. Specific heat data show the same $T^2$ dependence below $T_f$ that has been seen in SCGO(x), suggesting that this behaviour is a universal property for this class of 2D frustrated antiferromagnets.

Jarosites are a family of minerals of general formula $AFe_3(SO_4)_2(OH)_6$ (where A = $Na^+$, $K^+$, $Rb^+$, $Ag^+$, $Tl^+$, $NH_4^+$, $H_3O^+$, $1/2Pb^{2+}$ or $1/2Hg^{2+}$)[10]. They have already been identified as kagomé antiferromagnets, in which $FeO_6$ octahedra are linked through their vertices in layers that are well-separated by hydroxide, sulphate and hydronium groups. The magnetic ion $Fe^{3+}$ has a $^6A_{1g}$ ground term which will approximate very well to a S=5/2 Heisenberg spin. However, until recently[11] it has not been recognised by those studying their magnetic properties that jarosites are prone to nonstoichiometry; a significant proportion of $A^+$ ions are replaced by $H_3O^+$, and $Fe^{3+}$ is lost with the accompanying protonation of the hydroxyl groups. This results in a kagomé lattice with a typical coverage of 83 - 95%. An exception to this is the hydronium salt which can be prepared with a coverage of approximately 97%[11].

Partially deuterated hydronium jarosite was prepared by a method based on an established hydrothermal route[12]. 40g of $Fe_2(SO_4)_3 \cdot nH_2O$ was dissolved in $D_2O$ and made up to 300ml then heated in a PTFE-lined stainless steel bomb at 140°C for 4 hours. The deuteronium jarosite precipitate was washed with $D_2O$ and dried at 120°C. A new material, $(H_3O)Ga_3(OH)_6(SO_4)_2$, was prepared as a diamagnetic analogue for subtracting the phonon contribution from the specific heat measurements. 2g of $Ga_2(SO_4)_3$ was made up to $5cm^3$ in deionised water and heated at 140°C for 17 hours in a sealed Pyrex tube. The resulting precipitate was washed and dried at 120°C then characterised structurally by powder X-ray diffraction using a Rigaku "Geigerflex" diffractometer with CuK$\alpha$ radiation. The structure was refined from these data using GSAS[13] from the $R\bar{3}m$ jarosite structure[11] with lattice parameters a = 7.1754(23) Å and c = 17.1635(6) Å.

Measurements of $\chi_{dc}$ were carried out between 1.8K and 330K using a Quantum Design MPMS$_2$ SQUID Magnetometer. Fig. 2(a) clearly shows the onset of spin glass-like behaviour at $T_f$ = 13.8K as a divergence between zero-field-cooled (ZFC) and field-cooled (FC) measurements, where the cooling and the applied field were both 100G. The anomaly at 4.5K is believed to be an artefact of the temperature control; it is not seen in the measurements of $\chi_{ac}$ or specific heat which showed no anomalies other than that at $T_f$ down to 15mK. At higher temperatures the susceptibility conforms approximately to Curie-Weiss behaviour, with some curvature in the dependence of $1/\chi_{dc}$ on T between 150 and 300K as a result of short-range 2D correlations. Extrapolation of this curve to $1/\chi_{dc}=0$ provides an estimate for $\theta_{CW}$ of –1500±300K. Short-range 2D antiferromagnetic correlations are also believed to be responsible for reducing the effective moment at 300K from the spin-only value of $5.9\mu_B$ to approximately $4.3\mu_B$. Both $T_f$ and $\theta_{CW}$ differ slightly for the hydronium and deuterated forms; $T_f$ has also been found to vary between samples of the same material, ranging from 12 to 18K for samples of the hydronium salt of the same nominal composition[11]. Small variations in preparative conditions lead to small changes in lattice coverage, and the limits of the precision of our chemical analysis will encompass a range of values of $T_f$ and $\theta_{CW}$. Jarosites in which hydronium is exchanged for $K^+$, $Tl^+$ or $NH_4^+$ show long-range magnetic ordering transitions at ~ 50K[10]. The structures of all these jarosites are very similar and it is unlikely that the inter- or intraplane exchange pathways differ significantly. What *is* different is the degree of lattice coverage: the salts that order have relatively low coverage (<90%) while the hydronium salts have a coverage greater than 90%. Thus, it appears that *lower* coverage is found to lead to a *lower* value of $\theta_{CW}$ but a *higher* value of $T_f$, consistent with what is observed in SCGO(x). There may also be small differences in single-ion anisotropy and interplane exchange between the hydronium and deuterated salts.

The relative values of $T_f$ and $\theta_{CW}$ for a given sample are broadly consistent with theoretical predictions: Monte Carlo simulations on a Heisenberg kagomé antiferromagnet with a small XY exchange anisotropy ($J_x=J_y > J_z$) indicate that the flipping of single stars freezes out at $\approx \theta_{CW}/96$ ($\approx$ 15K for deuteronium iron jarosite), but the simulations were not performed for sufficiently long to study the freezing of longer closed folds[2]. When an Ising anisotropy is present, freezing is expected at $T = 0.058J$ ($\approx$ 4K for deuteronium iron jarosite) accompanied by a parasitic ferromagnetic moment[3].

In-phase and out-of-phase components of the $\chi_{ac}$ were measured over the temperature range 5-25K at frequencies of 1.157 - 1157 Hz using a standard mutual inductance technique. The data are presented in Fig. 2(b), showing the fall of $T_f$ with measuring frequency $\omega$ in a manner typical of an 'ideal' spin glass: $\Delta T_f/T_f \Delta \log \omega = 0.010$, the value also found for *Au*Fe[14]. These data were fitted to the empirical Vogel-Fulcher law $\omega = \omega_0 \exp[-E_a/(k(T_f-T_0))]$ with the activation energy $E_a = 64$K, the ideal glass temperature $T_0 = 15.6$K and $\tau_0 = 1/\omega_0 = 1.6 \times 10^{-12}$ sec. Specific heat data were taken for both $(D_3O)Fe_3(SO_4)_2(OD)_6$ and $(H_3O)Ga_3(SO_4)_2(OH)_6$ over the temperature range 0.4-40K. The data for the iron compound were corrected for the phonon background measured on the gallium compound (which contributes about 10% of the signal at these temperatures) and are displayed as a graph of $C_m(T)/T$ against T in Fig. 2(c) to bring out the $T^2$ dependence below $T_f$. This contrasts with conventional spin glasses where localised modes with an energy-independent density of states dominate at low temperatures, giving a linear dependence of $C_m(T)$ on T. A $T^2$ dependence implies propagating spin-waves based on long-range magnetic order in a conventional 2D antiferromagnet, or from the frozen spin state in a 2D spin-glass[15]; similar behaviour has been observed in SCGO(x). Deuteronium jarosite also ressembles SCGO(x) in the small fraction of the magnetic entropy released at $T_f$, being 18% of the full value of Rln(2S+1) for free spins S compared with approximately 7% at $T_f$ for SCGO(x). However, in contrast to SCGO(x) and also to 'ideal' spin glasses, no peak is observed in $C_m(T)/T$; instead, $C_m(T)/T$ changes monotonically with T over a wide range of temperature.

Neutron powder diffraction measurements were performed on deuteronium jarosite to probe the magnetic pair correlation function more incisively. Diffraction patterns were taken on the multidetector diffractometer C2 at the NRU Reactor, Chalk River at a wavelength of 1.5041Å. Data taken at 2K were refined successfully from the $R\bar{3}m$ jarosite structure[11] giving a = 7.3445(7)Å and c = 16.9037(16)Å. No additional Bragg peaks were seen at the lowest temperature; instead, we observed a broad asymmetric peak centred at Q ≈ 1.05Å$^{-1}$ indicative of short-range magnetic correlations (Fig. 3). Nuclear Bragg peaks were excluded from the diffuse scattering, and the data that remained were fitted to a Warren function[16] which gives the scattering power $P_{2\theta}$ of a two-dimensional disordered structure as:

$$P_{2\theta} = Km \frac{F_{hk}^2(1+\cos^2 2\theta)}{2(\sin\theta)^{3/2}} \left(\frac{L}{\sqrt{\pi}\lambda}\right)^{1/2} F(a)$$

where $a = (2\sqrt{\pi}L/\lambda)(\sin\theta - \sin\theta_0)$, K is a constant, m is the multiplicity of the reflection, $F_{hk}$ is the two-dimensional structure factor, λ is the wavelength, L is a two-dimensional correlation length, $\theta_0$ is the peak position, and the function $F(a)$, which is proportional to an integral of the scattering strength over different sizes and orientations of correlated moments, is calculated self-consistently from $L$ and the scattering geometry[16]. This yielded a two-spin correlation length of 19±2Å, the approximate point to point distance across an individual kagomé star. The fitted peak is centred at $Q = 1.05$Å$^{-1}$, close to the position (1.09Å$^{-1}$) of the (2 1) peak of the $\sqrt{3} \times \sqrt{3}$ spin structure. Comparison with the pattern obtained at 25K revealed no significant change, indicating that short-range ordering is largely established before the 13.8K transition in $\chi_{dc}$.

The dynamic character of the short-range magnetic correlations was studied in further detail with the triple axis diffractometer N5 at the NRU Reactor, initially configured as a two-axis instrument with an incident energy of 5THz ($\lambda = 1.989$Å) and a Si(111) monochromator to reduce contamination of the beam by neutrons of wavelength $\lambda/2$. $S(Q)$ measured at $Q = 1.4$ Å$^{-1}$, which is in the region of the strong diffuse scattering but avoids the nearby Bragg peak, rises steadily as the sample is cooled from approximately $4T_f$. The development of short-range magnetic correlations was more marked when the incident neutron energy was reduced to 3THz ($\lambda = 2.568$Å) and the instrument was reconfigured as a triple axis spectrometer with an energy resolution of 0.22THz (Fig. 4). After normalisation to the intensity of the (003) reflection, the scattering intensity at this wavevector was found to be reduced by a factor of 4 by the change in spectrometer configuration. Thus, 75% of the neutrons scattered magnetically at 2K are from fluctuations faster than about 0.2THz. The observation that these correlations set in at temperatures higher than the freezing tempetaures observed with DC and AC susceptibility is consistent with the much higher frequency probed in these relatively low resolution measurements.

In conclusion, we have reported measurements of the static and dynamic magnetic correlations in deuteronium jarosite, $(D_3O)Fe_3(SO_4)_2(OD)_6$, which presents a highly covered (97±1%), magnetically isolated, kagomé lattice of antiferromagnetically coupled Heisenberg spins S=5/2. Our results indicate that short-range $\sqrt{3} \times \sqrt{3}$ correlations coexist with propagating, spin-wave like modes as well as behaviour more typical of a spin-glass such as FC-ZFC irreversibility in $\chi_{dc}$ and a frequency-dependent $T_f$. Similar behaviour is seen in SCGO(x). This suggests that this a universal property of this class of 2D frustrated magnets, as well as a feature which should be reproduced in theoretical work. It is interesting to note that these effects are seen in a relatively classical, S=5/2 system, suggesting that quantum effects are not significant. Our observations are consistent with a system of interacting dynamical extended spin defects in a manifold of spin states derived from the $\sqrt{3} \times \sqrt{3}$ structure. More theoretical work needs to be done on the energy and dynamics of collections of such defects. In particular, we need a clearer understanding of the energy scale of the defect-defect interactions relative to J, and of the influence of quenched diamagnetic defects on the spin correlations.

The authors would like to thank M.F. Collins for introducing them to jarosites as kagomé magnets, J.E. Greedan for drawing their attention to the Warren function and supplying programs to fit it, and S.T. Bramwell for enlightening conversations. We are also grateful to G.J. Nieuwenhuys for the use of the AC susceptibility and SQUID apparatus at the Kamerlingh Onnes Laboratory, Leiden University. We thank CIAR, EPSRC, NATO and NSERC for financial support, and the technical staff at Chalk River for their expertise in assisting with the neutron scattering measurements.

# Figure Captions

1. Spin arrangements and defects for a kagomé antiferromagnet : (a) the q = 0 structure supporting an open defect bounded by the dashed lines and (b) the $\sqrt{3}$ x $\sqrt{3}$ structure supporting a closed defect denoted by the bold hexagon.

2. Changes in (a) $\chi_{dc}$ , (b) $\chi'_{ac}$ and (c) $C_m$ in the region of $T_f$ for deuteronium jarosite. (a) shows ZFC (closed circles) and FC (open circles) data while the insert depicts ZFC $\chi_{dc}$ (circles) and $1/\chi_{dc}$ (triangles) over a wider temperature range. (b) shows the response of the sample with a driving frequency $\omega$ of 1.157Hz, while the insert shows how $T_f$ falls with $\omega$ in a geometric progression from 1.157 to 1157Hz. The straight line through the low-temperature data in (c) represents a $T^2$ dependence of $C_m$ on T.

3. Diffuse neutron scattering profile from deuteronium jarosite at 2K with a fit to the Warren lineshape function centred near the (2 1) reflection of the $\sqrt{3}$ x $\sqrt{3}$ spin structure. Breaks in the data and fit indicate regions where nuclear Bragg peaks have been subtracted.

4. Temperature dependence of the diffuse scattering from deuteronium jarosite sampled at Q = 1.4Å$^{-1}$ and $\Delta\omega \leq 0.22$ THz. The line through the data provides a guide to the eye.

# References

[1] P. Chandra and P. Coleman, Phys. Rev. Lett. **66**, 100 (1991); A. B. Harris, C. Kallin and A. J. Berlinsky, Phys. Rev. B **45**, 2899 (1992); A. Chubukov, Phys. Rev. Lett. **69**, 832 (1992); J. T. Chalker, P. C. W. Holdsworth and E. F. Shender, Phys. Rev. Lett. **68**, 855 (1992); J. N. Reimers and A. J. Berlinsky, Phys. Rev. B **48**, 9539 (1993); H. Asakawa and M. Suzuki, Int. J. Mod. Phys. B **9**, 933 (1995)

[2] I. Ritchey, P. Chandra and P. Coleman, Phys. Rev. B **47**, 15342 (1993); P. Chandra, P. Coleman and I. Ritchey, J. Phys. I **3**, 591 (1993)

[3] S. T. Bramwell and M. J. P. Gingras, Unpublished work (1994)

[4] X. Obradors *et al*, Solid State Commun. **65**, 189 (1988); A. P. Ramirez, G. P. Espinosa and A. S. Cooper, Phys. Rev. Lett. **64**, 2070 (1990)

[5] S.-H. Lee *et al,* Phys. Rev. Lett. **76**, 4424 (1996)

[6] Y. J. Uemura *et al*, Phys. Rev. Lett. **73**, 3306 (1994)

[7] C. Broholm *et al*, Phys. Rev. Lett. **65**, 3173 (1990)

[8] A. P. Ramirez, G. P. Espinosa and A. S. Cooper, Phys. Rev. B **45**, 2505 (1992)

[9] A. Harrison and T. E. Mason, J. Appl. Phys. **67**, 5424 (1990)

[10] J. E. Dutrizac and S. Kaiman, Can. Miner. **14**, 151 (1975); M. G. Townsend, G. Longworth and E. Roudaut, Phys. Rev. B **33**, 4919 (1986); C. Broholm *et al*, Unpublished work (1992); M. Takano *et al*, J. Phys. Soc. Japan **25**, 902 (1968)

[11] A. S. Wills and A. Harrison, Faraday Trans. **92**, 2161 (1996)

[12] J. Kubisz, Miner. Pol. **1**, 45 (1970)

[13] *Generalised Structure Analysis System*, A. C. Larson and R. B. Von Dreele (1995), Obtainable from the authors at LANSCE, MS-H805, Los Alamos National Laboratory, NM 87454, USA

[14] J. A. Mydosh *Spin Glasses,* , Taylor and Francis, London, 1993.

[15] B. I. Halperin and W. M. Saslow, Phys. Rev. **B 16**, 2154 (1977)

[16] B. E. Warren, Phys. Rev. **59**, 693 (1941); J. E. Greedan *et al*, J. Solid State Chem. **116**, 118 (1995)

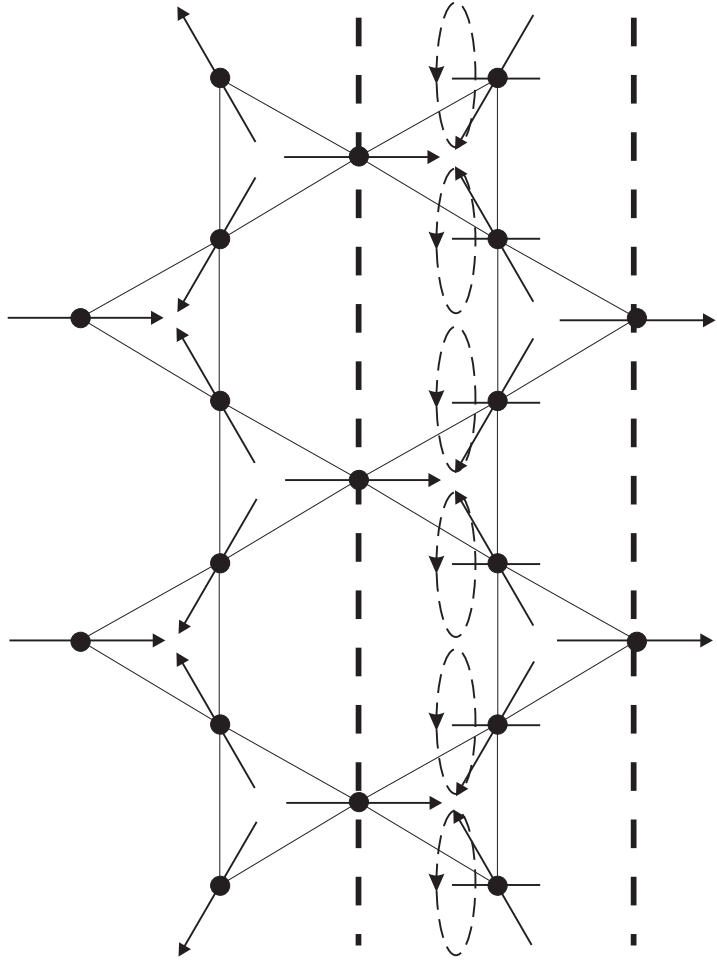

(a)

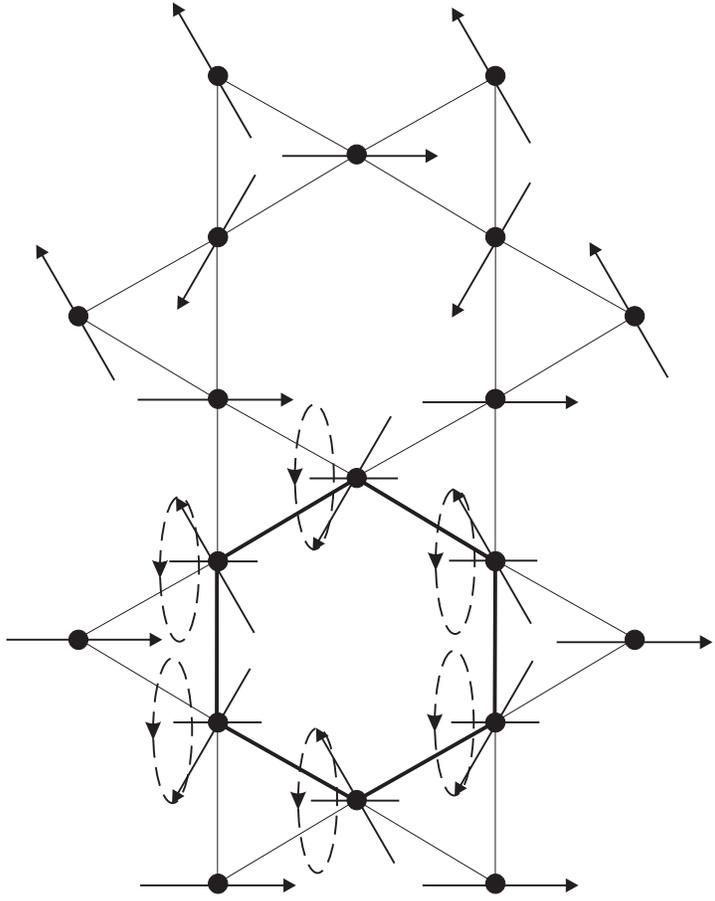

(b)

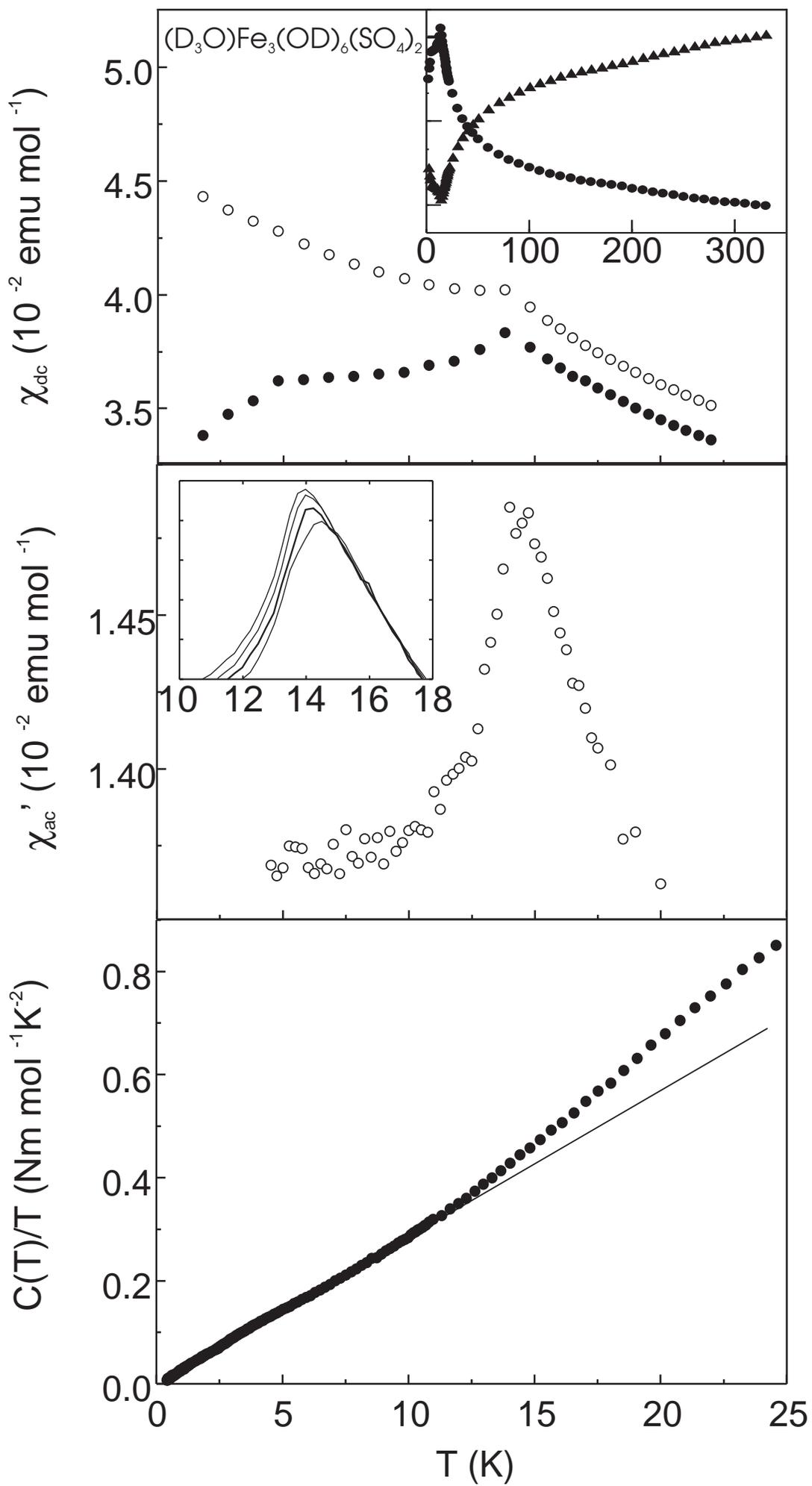

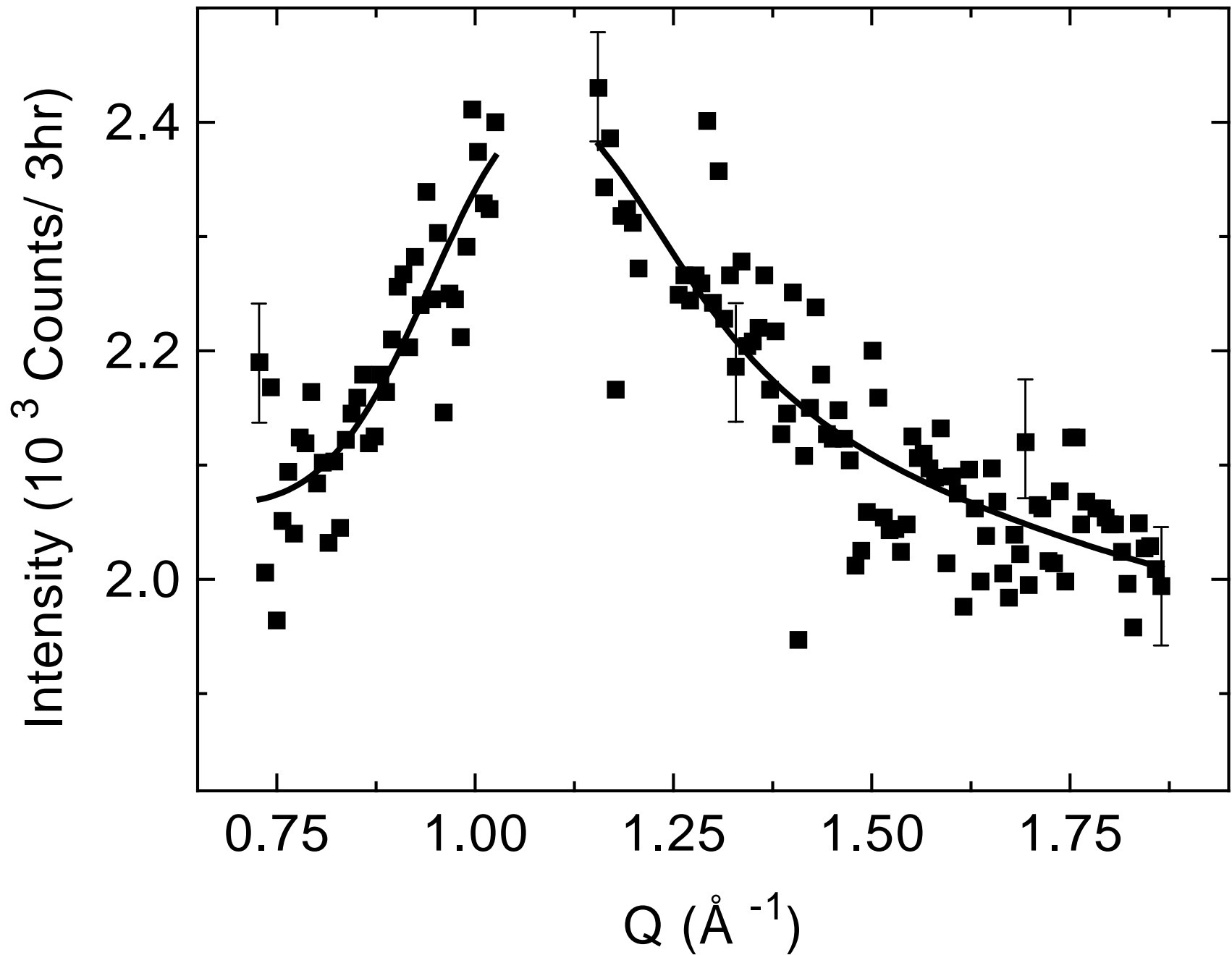

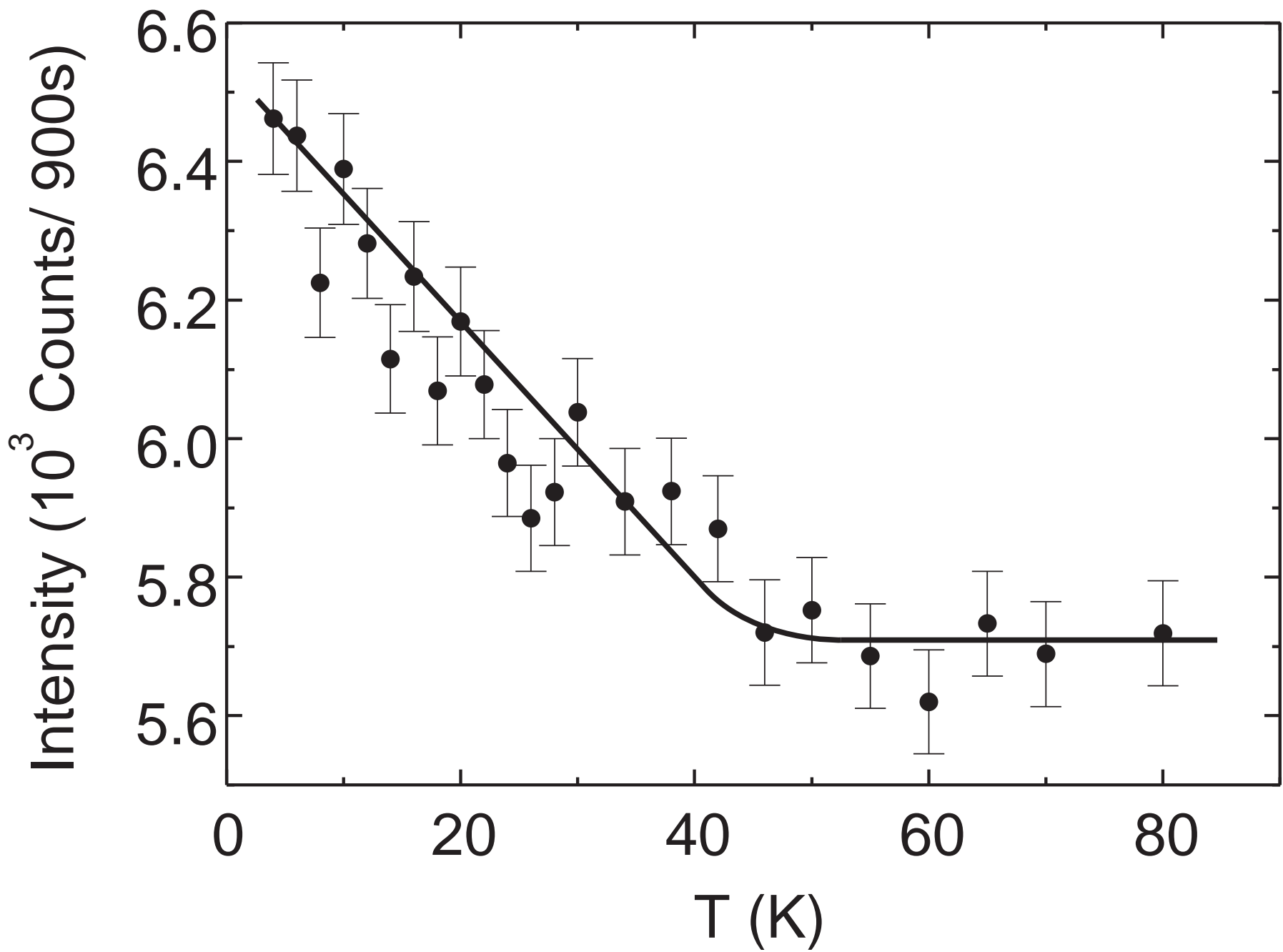